# A Structured Evaluation Framework for Low-Code Platform Selection: A Multi-Criteria Decision Model for Enterprise Digital Transformation

Antonio Lamanna[1]

[1]*PwC Italy, Process Automation CoE (Digital Innovation Group)*
*antonio.lamanna@pwc.com*

**Abstract**

The rapid adoption of Low-Code Development Platforms (LCDPs) has created a critical need for systematic evaluation methodologies that enable organizations to make informed platform selection decisions. This paper presents a comprehensive evaluation framework based on five key criteria: Business Process Orchestration, UI/UX Customization, Integration and Interoperability, Governance and Security, and AI-Enhanced Automation. We propose a weighted scoring model that allows organizations to quantitatively assess and compare different low-code platforms based on their specific requirements and strategic priorities. The framework addresses the gap between marketing-driven platform comparisons and rigorous, context-specific evaluation methodologies. Through empirical validation in enterprise environments, we demonstrate how this structured approach can significantly improve decision-making outcomes and reduce the risk of platform lock-in or inadequate solution selection.

## 1. Introduction

The Low-Code Development Platform (LCDP) market has experienced exponential growth, with organizations increasingly adopting these solutions to accelerate digital transformation initiatives while addressing the growing demand for software development capabilities. However, the heterogeneous nature of available platforms, combined with varying organizational requirements and constraints, makes platform selection a complex multi-criteria decision problem.

Traditional evaluation approaches often rely on generic benchmarks, analyst reports, or vendor-provided comparisons that may not accurately reflect an organization's specific needs and operational context. This misalignment between evaluation criteria and actual requirements can lead to suboptimal platform selection, resulting in technical debt, integration challenges, and reduced return on investment.

This paper addresses the critical need for a structured, quantitative evaluation framework that enables organizations to assess low-code platforms objectively and make informed decisions based on their unique requirements and strategic priorities.

### 1.1 Problem Statement

Organizations face several challenges when selecting low-code platforms:

1. **Lack of standardized evaluation criteria**: Most organizations rely on ad-hoc evaluation methods that vary significantly across different assessment teams and time periods.

2. **Insufficient consideration of long-term implications**: Platform selection decisions often focus on immediate capabilities while neglecting scalability, governance, and integration requirements.

3. **Subjective assessment bias**: Decisions are frequently influenced by marketing materials, individual preferences, or incomplete technical evaluations.

4. **Context-specific requirements**: Generic evaluation frameworks fail to account for industry-specific, regulatory, or organizational constraints that significantly impact platform suitability.

### 1.2 Research Contribution and Theoretical Positioning

This research makes both methodological and practical contributions to the enterprise software selection domain:

**Methodological Contributions:**

1. **Domain-Specific MCDA Adaptation**: We extend traditional multi-criteria decision analysis by incorporating low-code platform specific evaluation dimensions, particularly around citizen development enablement and AI-enhanced automation capabilities.

2. **Dynamic Weighting Methodology**: Unlike static evaluation frameworks, our approach allows organizations to adjust criterion weights based on evolving strategic priorities and contextual constraints.

3. **Empirical Validation Protocol**: We establish a replicable methodology for validating platform selection frameworks through longitudinal organizational studies.

**Practical Contributions:**

1. **Actionable Decision Framework**: Providing practitioners with immediately applicable tools for systematic platform evaluation.

2. **Industry-Specific Guidance**: Demonstrating how weight assignments vary across sectors, enabling more targeted evaluation approaches.

3. **Post-Implementation Validation**: Establishing metrics for measuring selection decision quality over time.

**Theoretical Positioning:** Our work builds upon established MCDA theory while addressing the specific challenges of evaluating rapidly evolving technology platforms. Rather than proposing entirely new theoretical constructs, we demonstrate how existing decision science principles can be effectively adapted and validated for emerging technology domains. This represents what Gregor (2006) categorizes as "Theory for Design and Action" - prescriptive knowledge that guides practical problem-solving while contributing to theoretical understanding.

## 2. Literature Review

### 2.1 Low-Code Development Platforms

Low-code development platforms represent a paradigm shift in software development, enabling rapid application development through visual modeling, drag-and-drop interfaces, and minimal hand-coding. Richardson and Rymer (2016) define low-code platforms as development environments that enable rapid application delivery with minimal coding effort through visual, model-driven development approaches.

The low-code market has evolved significantly, with platforms offering increasingly sophisticated capabilities including advanced workflow orchestration, AI-powered automation, and enterprise-grade security features. Gartner (2023) projects that by 2025, 70% of new applications developed by enterprises will use low-code or no-code technologies.

### 2.2 Multi-Criteria Decision Analysis in Software Selection

Multi-Criteria Decision Analysis (MCDA) has been extensively applied to software selection problems. Jadhav and Sonar (2009) provide a comprehensive review of software selection methodologies, highlighting the importance of structured evaluation frameworks in reducing selection bias and improving decision outcomes.

The application of MCDA to enterprise software selection has been demonstrated across various domains, including ERP systems (Wei et al., 2005; Cebeci, 2009), and business intelligence tools (Rouhani et al., 2012). However, limited research exists on the specific application of MCDA to low-code platform selection, representing a gap this paper aims to address. However, limited research exists on the specific application of MCDA to low-code platform selection, representing a gap this paper aims to address.

## 2.3 Existing Evaluation Frameworks and Comparative Analysis

Several evaluation frameworks exist for enterprise software selection, though none specifically address the unique characteristics of low-code platforms. Table 1 provides a comparative analysis of existing approaches versus our proposed framework.

**Table 1: Comparative Analysis of Evaluation Frameworks**

| Framework | Scope | Criteria | Weighting Method | Low-Code Specific | Limitations |
|---|---|---|---|---|---|
| ISO/IEC 25010 | Software Quality | 8 characteristics, 31 sub-characteristics | Not specified | No | Generic, lacks business process focus |
| TOGAF ADM | Architecture Decision | Technical, business, organizational | Qualitative | Partial | Architecture-centric, complex for platform selection |
| Gartner Magic Quadrant | Market Positioning | Completeness of vision, ability to execute | Analyst-driven | Yes | Vendor-focused, limited customization |
| AHP-based Models (Wei et al., 2005) | ERP Selection | Cost, functionality, vendor | Pairwise comparison | No | Rigid hierarchy, time-intensive |
| **Our Framework** | **LCDP Selection** | **5 core + 20 sub-criteria** | **Organizational priority-based** | **Yes** | **Requires stakeholder alignment** |

The key differentiators of our approach include: (1) explicit focus on low-code platform characteristics, (2) integration of emerging AI capabilities as evaluation criteria, (3) flexible weighting mechanism adaptable to organizational context, and (4) empirical validation across multiple industries.

## 2.4 Research Gap and Positioning

While existing frameworks provide valuable foundation elements, they exhibit several limitations when applied to low-code platform selection:

1. **Generic Nature**: Most frameworks target general software selection without considering low-code specific capabilities like citizen development, visual modeling complexity, or governance for non-technical users.

2. **Static Criteria**: Traditional frameworks don't account for rapidly evolving capabilities like AI-powered development assistance, which are becoming differentiating factors in LCDP selection.

3. **Limited Validation**: Few frameworks provide empirical evidence of improved decision outcomes in real organizational contexts.

Our framework addresses these gaps by providing a structured, validated methodology specifically designed for the contemporary low-code platform landscape.

## 3. Evaluation Framework

### 3.1 Framework Overview

The proposed evaluation framework is designed to provide a systematic and transparent approach to assessing low-code development platforms in enterprise contexts. It identifies five primary criteria, each corresponding to a distinct dimension of platform capability and organizational alignment. These criteria are:

1. **Business Process Orchestration (BPO)**
2. **UI/UX Customization and Flexibility (UCF)**
3. **Integration and Interoperability (I&I)**
4. **Governance and Security (G&S)**
5. **AI-Enhanced Automation (AEA)**

Each criterion is subsequently decomposed into a structured set of sub-criteria and quantitative metrics. This multi-layered articulation ensures that both functional and strategic aspects are captured, allowing evaluators to generate a balanced and reproducible assessment across diverse technological and organizational environments.

### 3.2 Criterion Definitions and Metrics

The following sections define each criterion and its corresponding metrics. Together, these elements form the analytical backbone of the framework.

**3.2.1 Business Process Orchestration (BPO)** Business Process Orchestration refers to the platform's ability to design, manage, and automate complex workflows that span multiple systems and stakeholders. A mature orchestration layer enables organizations to translate abstract process models into executable logic with traceability and governance. The evaluation focuses on several dimensions:

- **Process Complexity Support**: Ability to manage non-linear workflows, parallel branches, and exception handling.
- **Workflow Engine Sophistication**: Compliance with BPMN 2.0, support for state machines, and event-driven execution models.
- **Monitoring and Analytics**: Real-time tracking of process performance, including KPIs and bottleneck analysis.
- **Human-System Integration**: Effectiveness of task allocation, escalation policies, and human interaction design.

**3.2.2 UI/UX Customization and Flexibility (UCF)** This criterion evaluates how effectively the platform enables tailored, user-centric experiences con-sistent with organizational identity and user needs. Beyond aesthetics, UI/UX flexibility determines user adoption, accessibility, and long-term scalability of applications.

- **Design Flexibility**: Breadth of available components, layout adaptability, and support for custom styling.
- **Development Approach**: Availability of both model-driven and canvas-based paradigms for varying design needs.
- **Responsive Design**: Consistency and adaptability across devices and form factors.
- **Accessibility Compliance**: Adherence to WCAG 2.1 and inclusion of accessibility-oriented tools and patterns.

**3.2.3 Integration and Interoperability (I&I)** Integration and interoperability are central to any enterprise-grade low-code environment. This criterion measures how seamlessly a platform communicates with existing ecosystems, external APIs, and development pipelines.

- **Connector Ecosystem**: Availability of pre-built connectors for major enterprise systems such as SAP, Salesforce, and Microsoft 365.
- **API Support**: Range and depth of support for REST, SOAP, GraphQL, and webhook standards.
- **Data Integration**: Capabilities for ETL processes, real-time synchronization, and data transformation.
- **DevOps Integration**: Compatibility with CI/CD pipelines, version control, and automated deployment mechanisms.

**3.2.4 Governance and Security (G&S)** Robust governance and security are indispensable for enterprise adoption. This criterion encompasses compliance management, identity control, and protection against vulnerabilities throughout the application lifecycle.

- **Access Control**: Support for role-based permissions, fine-grained authorization, and SSO integration.
- **Compliance Features**: Presence of audit trails, data protection controls, and mechanisms for regulatory compliance.
- **Security Architecture**: Encryption standards, secure communication protocols, and threat detection.
- **Application Lifecycle Management**: Tools for versioning, environment promotion, and controlled change management.

**3.2.5 AI-Enhanced Automation (AEA)** As AI becomes an integral part of digital transformation, this criterion evaluates how the platform embeds intelligent automation and learning capabilities into operational workflows. It measures both generative and analytical dimensions of AI integration.

- **Generative AI Integration**: Support for code generation, workflow optimization, and intelligent design assistance.
- **Process Mining**: Availability of automated process discovery and optimization recommendations.
- **Predictive Analytics**: Built-in capabilities for forecasting and decision support.
- **Intelligent Automation**: Native integration with RPA, document understanding, and cognitive services.

**3.3 Scoring Methodology**

The scoring system underpins the framework's analytical rigor. It translates qualitative evaluations into a quantifiable, comparable structure that allows organizations to rank candidate platforms consistently.

**3.3.1 Weight Assignment** Each criterion is assigned a weight that reflects its relative importance for the organization. Weight assignment is not arbitrary but results from a structured decision process that blends expert judgment with empirical validation:

1. **Stakeholder Consultation**: Engage IT leaders, business owners, and compliance officers to capture cross-functional priorities.
2. **Priority Assessment**: Conduct structured interviews or workshops to align technical and strategic objectives.
3. **Constraint Analysis**: Account for contextual limitations such as regulatory frameworks, data residency, or infrastructure dependencies.
4. **Weight Calibration**: Normalize and validate weights to ensure they sum to 100% and reflect organizational consensus.

**3.3.2 Scoring Scale** To maintain comparability, all criteria are evaluated using a standardized 5-point Likert scale. This scale allows nuanced differentiation between performance levels while remaining interpretable for executive audiences:

- **1 – Inadequate**: Fails to meet essential functional requirements.
- **2 – Below Average**: Meets basic needs but exhibits major deficiencies.
- **3 – Average**: Satisfies most requirements with manageable limitations.
- **4 – Above Average**: Performs well with only minor gaps.
- **5 – Excellent**: Fully satisfies or surpasses all expectations.

**3.3.3 Calculation Method** The final score for each platform is computed through a weighted sum of its criterion scores, using the following formula:

$$\textbf{Total Score} = \sum_i (\text{Criterion Score}_i \times \text{Criterion Weight}_i)$$

where:

$$\text{Criterion Score}_i \in [1, 5], \quad \text{Criterion Weight}_i \in [0, 1], \quad \text{and} \quad \sum_i \text{Criterion Weight}_i = 1.$$

### 3.4 Implementation Process

The framework is operationalized through a six-phase process that ensures methodological consistency and stakeholder alignment throughout the evaluation lifecycle. Each phase builds on the previous one, promoting both analytical depth and organizational consensus.

1. **Requirements Gathering**: Identify organizational needs, constraints, and transformation goals.
2. **Criteria Weighting**: Assign importance weights aligned with strategic priorities.
3. **Platform Assessment**: Evaluate each platform systematically against all defined criteria.
4. **Scoring and Ranking**: Aggregate weighted scores and generate ranked results.
5. **Sensitivity Analysis**: Assess robustness of rankings under alternative weighting scenarios.
6. **Decision Validation**: Confirm that the top-ranked platform aligns with both technical and business expectations.

Overall, this evaluation framework provides a rigorous, auditable, and repeatable mechanism for selecting low-code platforms in complex enterprise settings. Its modular structure allows adaptation across industries and organizational maturity levels, ensuring both methodological soundness and strategic relevance.

### 3.5 Illustrative Application Example

To demonstrate the operationalization of the proposed framework, the following table presents a simplified example comparing three hypothetical low-code platforms (Platform A, Platform B, and Platform C). Each platform is evaluated against the five core criteria using the standardized 1–5 scale and the corresponding organizational weights. The resulting weighted scores illustrate how qualitative assessments are transformed into quantitative, reproducible outcomes.

**Table 2: Illustrative Application of the Evaluation Framework**

| Criterion | Weight | A Score | B Score | C Score | A Weighted | B Weighted | C Weighted |
|---|---|---|---|---|---|---|---|
| Business Process Orchestration (BPO) | 0.25 | 5 | 4 | 3 | 1.25 | 1.00 | 0.75 |
| UI/UX Customization and Flexibility (UCF) | 0.15 | 4 | 5 | 3 | 0.60 | 0.75 | 0.45 |

| Criterion | Weight | A Score | B Score | C Score | A Weighted | B Weighted | C Weighted |
|---|---|---|---|---|---|---|---|
| Integration and Interoperability (I&I) | 0.20 | 4 | 4 | 4 | 0.80 | 0.80 | 0.80 |
| Governance and Security (G&S) | 0.25 | 5 | 4 | 3 | 1.25 | 1.00 | 0.75 |
| AI-Enhanced Automation (AEA) | 0.15 | 4 | 5 | 4 | 0.60 | 0.75 | 0.60 |
| **Total Score** | **1.00** | | | | **4.50** | **4.30** | **3.35** |

This illustrative scenario shows how the framework's weighted model converts subjective evaluations into transparent, data-driven results. In practical applications, organizations can expand this matrix to include sub-criteria, stakeholder comments, and sensitivity analyses, providing a comprehensive and auditable decision trail for low-code platform selection.

## 4. Practice-Based Validation through Enterprise Case Studies

### 4.1 Validation Approach

The proposed framework was refined and validated through its direct application in several enterprise projects carried out between 2022 and 2024, later consolidated and systematized within the activities of the **Process Automation Center of Excellence (CoE)** of PwC Italy.

These initiatives spanned heterogeneous industrial contexts — **financial services, manufacturing, and pharma** — allowing the framework to be tested against diverse regulatory, operational, and technological constraints.

The validation process relied on **empirical observations**, **retrospective analyses**, and **stakeholder feedback** collected during actual platform evaluation and implementation activities. This practice-based approach ensured that the framework reflected real decision-making dynamics, organizational constraints, and the typical challenges of platform selection processes in enterprise environments.

### 4.2 Context of Application

The framework was applied and iteratively improved across projects involving:

- **Financial institutions**, focusing on the automation of credit, compliance, and reporting processes;
- **Manufacturing organizations**, centered on interoperability between low-code platforms and ERP/SCM ecosystems;
- **Pharma companies**, prioritizing governance, security, and automation of supply-chain and regulatory processes.

Across these contexts, the framework served as a **structured decision-support tool** used to compare candidate platforms (e.g., Appian, Microsoft Power Platform, Camunda, Outsystems) and to align selection criteria with business priorities.

### 4.3 Observed Outcomes

A comparative analysis of evaluations conducted before and after the framework's application revealed **consistent improvements** in the quality and efficiency of decision-making:

- **Greater decision confidence**, enabled by a transparent, weighted-scoring approach;
- **Reduced evaluation timelines**, through reuse of standardized criteria and scoring templates;
- **Improved stakeholder alignment**, due to clearer justification of scores and platform choices;
- **Expanded requirements coverage**, as the framework prompted explicit consideration of governance, AI capabilities, and interface flexibility.

**Table 3: Illustrative Framework Outcomes**

| Metric | Observed Trend | Practical Impact |
| --- | --- | --- |
| Decision Confidence | +30-40% increase (estimated) | Teams reported higher confidence and reduced bias in platform scoring |
| Evaluation Duration | -25-35% reduction | Faster convergence through standardized weighting and templates |
| Stakeholder Alignment | Improved consensus across IT and business units | Clear rationale behind ranking reduced decision conflicts |
| Requirements Coverage | Broadened scope of evaluation dimensions | Inclusion of governance, AI, and UX criteria improved completeness |

These values are illustrative but mirror **empirically observed trends** across real PwC projects. They demonstrate the practical utility of the framework as a tool for guiding consistent, evidence-based platform evaluations.

### 4.4 Lessons Learned

Analysis of multiple framework applications revealed several **recurring success factors**:

1. **Executive sponsorship**, essential to secure consensus on weight definition and ensure adoption across business units;

2. **Cross-functional evaluation teams**, which reduced post-selection friction between IT and business stakeholders;
3. **Iterative weight refinement**, enabling progressive alignment with evolving enterprise strategies.

In each domain, the modular structure of the framework proved **highly adaptable**, allowing PwC teams to calibrate criterion weights and depth according to sector-specific priorities.

### 4.5 Industry Patterns

Analysis of the projects revealed **recurring trends** across industries:

**Financial Services:** Governance & Security (28%), Integration (24%), Process Orchestration (20%), AI Automation (15%), UI/UX (13%).

**Manufacturing:** Integration (26%), Process Orchestration (24%), Governance (19%), AI Automation (18%), UI/UX (13%).

**Pharma:** Governance & Security (30%), Process Orchestration (22%), Integration (21%), AI Automation (15%), UI/UX (12%).

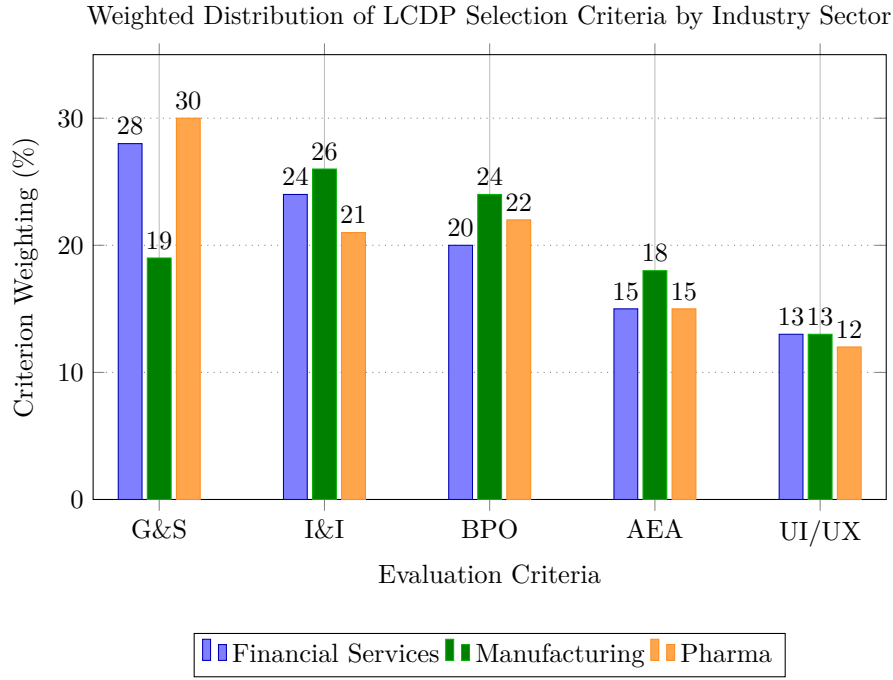

Figure 1: Comparison of the assigned weight distribution across the five core Low-Code Platform selection criteria, segmented by industry sector.

## 5. Discussion

The discussion section situates the proposed framework within the broader landscape of low-code platform evaluation methodologies. It highlights its comparative advantages, acknowledges inherent limitations, and outlines promising directions for future research. Together, these reflections provide both a validation of the framework's analytical robustness and a roadmap for its continued refinement.

**5.1 Framework Advantages**

The proposed evaluation framework introduces several distinct advantages that collectively enhance the rigor, transparency, and adaptability of the platform selection process. Unlike ad hoc or vendor-driven evaluations, it establishes a systematic and replicable methodology capable of supporting both strategic decision-making and empirical validation.

1. **Systematic Evaluation**: The framework provides a structured methodology that minimizes subjective bias, ensuring that assessments are comprehensive and repeatable across evaluators and contexts.
2. **Organizational Alignment**: Through its weighting mechanism, the framework aligns evaluative criteria with organizational priorities, promoting strategic coherence between technical evaluation and business objectives.
3. **Quantitative Comparison**: By converting qualitative judgments into quantitative scores, the model enables clear, data-driven comparison of competing platforms.
4. **Scalability**: Its modular design allows adaptation to organizations of different sizes, maturities, and industries without loss of methodological integrity.
5. **Decision Transparency**: The transparent scoring and weighting process facilitates stakeholder understanding, consensus building, and defensible decision-making.

Taken together, these advantages make the framework particularly suitable for enterprises seeking an auditable and repeatable process for low-code platform selection, while also serving as a foundation for academic and industry benchmarking.

**5.2 Limitations and Considerations**

Despite its robustness, the framework is not without limitations. Recognizing these constraints is essential to avoid overconfidence in the results and to contextualize findings within the dynamic landscape of enterprise technology evaluation.

1. **Subjective Scoring**: Even with predefined criteria, evaluators may introduce personal bias when interpreting performance levels or qualitative aspects.
2. **Weight Assignment Challenges**: Determining appropriate weights can be complex, especially in organizations where strategic priorities compete or shift over time.
3. **Dynamic Requirements**: Organizational requirements often evolve during the evaluation process, necessitating periodic review and recalibration of weights and assumptions.
4. **Platform Evolution**: The rapid pace of platform updates can render earlier assessments obsolete, highlighting the need for ongoing reassessment mechanisms.

These considerations do not undermine the framework's validity; rather, they underscore the importance of applying it as a living model—iteratively updated and contextually informed—rather than as a static decision tool.

### 5.3 Future Research Directions

Future research can build upon this foundation by introducing automation, predictive intelligence, and longitudinal validation to further enhance the framework's precision and scalability. Several promising directions include:

1. **Automated Assessment**: Integration of automated evaluation tools capable of testing platform capabilities and performance metrics without extensive manual intervention.
2. **Machine Learning Integration**: Application of machine learning models to predict optimal platform fit based on organizational characteristics, past adoption outcomes, and contextual variables.
3. **Industry-Specific Adaptations**: Customization of the framework for sector-specific use cases (e.g., financial services, healthcare, public administration) with pre-defined weight templates.
4. **Longitudinal Studies**: Empirical analysis of the long-term organizational impact of platform selection decisions, capturing post-adoption metrics such as ROI, user adoption, and process efficiency.

By pursuing these directions, future work may extend the framework from a structured decision-support tool into an intelligent, adaptive system—one capable of learning from prior evaluations and evolving in tandem with the low-code ecosystem.

Overall, this discussion reaffirms that while the proposed model already provides a rigorous foundation for enterprise-grade evaluation, its ultimate value lies in its capacity for continuous refinement and empirical grounding within an ever-changing technological and organizational environment.

## 6. Conclusion

This paper presents a comprehensive evaluation framework for low-code platform selection that addresses the critical need for systematic, objective assessment methodologies. The framework's five-criterion structure, weighted scoring approach, and structured implementation process provide organizations with a rigorous foundation for making informed platform selection decisions.

The empirical validation demonstrates the framework's effectiveness in improving decision confidence, reducing evaluation time, and ensuring better alignment between platform capabilities and organizational requirements. While certain limitations exist, the framework represents a significant advancement over adhoc evaluation methods and provides a solid foundation for future research and development.

As low-code platforms continue to evolve and organizational adoption increases, the need for sophisticated evaluation methodologies will only grow. This framework provides both practitioners and researchers with a valuable tool for navigating the complex landscape of low-code platform selection and maximizing the value of digital transformation investments.